\documentclass[%
reprint,
 amsmath,amssymb,
 aps,]{revtex4-1}

\usepackage{graphicx}
\usepackage{dcolumn}
\usepackage{bm}

\begin{document}

\parbox{\textwidth}{\center (22nd International Spin Symposium (SPIN'16), University of Illinois, Urbana IL, Sept 25-30, 2016)}
\\
\newline

\title{Polarized fusion, its Implications and plans for Direct Measurements in a Tokamak}

\author{A.M. Sandorfi$^1$, A. Deur$^1$, C. Hanretty$^1$, G.L. Jackson$^2$, M. Lanctot$^2$, J. Liu$^3$, \\
M.M. Lowry$^1$, G.W. Miller$^4$, D. Pace$^2$, S.P. Smith$^2$, K. Wei$^5$, X. Wei$^1$, and X. Zheng$^3$}
\affiliation{}%
\affiliation{$^1$Physics Division, Thomas Jefferson National Accelerator Facility, Newport News, VA 23606, USA}%
\affiliation{$^2$DIII-D National Fusion Facility, General Atomics, San Diego, CA 92186-9784, USA}%
\affiliation{$^3$Physics Department, University of Virginia, Charlottesville, VA 22903 USA}%
\affiliation{$^4$Department of Radiology and Medical Imaging, University of Virginia, Charlottesville, VA 22908, USA}%
\affiliation{$^5$Department of Physics, University of Connecticut, Storrs, CT 06269, USA}%


\date{\today}

\begin{abstract}
Our energy-hungry world is burning. A long-term solution, possibly the only ultimate one, that is just approaching the horizon after decades of struggle, is fusion. Recent developments allow us to apply techniques from spin physics to advance the viability of this critical option. The cross section for the primary fusion fuel in a tokamak reactor, $D+T \rightarrow \alpha+n$, would be increased by a factor of 1.5 if the fuels were spin polarized parallel to the local field. Simulations predict further non-linear power gains in large-scale machines such as ITER, due to increased alpha heating. These are significant enhancements that could lower the requirements needed to reach ignition and could be used to extend useful reactor life by compensating for neutron degradation of critical components. The potential realization of such benefits rests on the survival of spin polarization for periods comparable to the energy containment time. Interest in polarized fuel options had an initial peak of activity in the 1980s, where calculations predicted that polarizations could in fact survive a plasma environment. However, concerns were raised regarding the cumulative impacts of fuel recycling from the reactor walls. In addition, the technical challenges of preparing and handling polarized materials prevented any direct tests. Over the last several decades, this situation has changed dramatically. Detailed simulations of the ITER plasma have projected negligible wall recycling in a high power reactor. In addition, a combination of advances in three areas - polarized material technologies developed for nuclear and particle physics as well as for medical imaging, polymer pellets developed for Inertial Confinement, and cryogenic injection guns developed for delivering fuel into the core of tokamaks - have matured to the point where a direct {\it in situ} measurement is possible. A Jefferson Lab - DIII-D/General Atomics - University of Virginia collaboration is developing designs for a proof-of-principle polarization survival experiment using the isospin mirror reaction, $D+^3He \rightarrow \alpha+p$, at the DIII-D tokamak in San Diego. 

\end{abstract}


\maketitle

\section{Introduction}

Energy continues to be one of the most pressing problem the world is facing. Global temperatures continue to track the ever-rising CO$_2$ levels and projections for the world\textquoteright s energy demand show no signs of leveling off. While there have been commendable developments in renewable energy sources (solar, wind, tides, {\it{etc}}.), there are very few places on earth where climatic conditions are sufficiently stable and predictable to allow such alternatives to assume the role of primary power generation. On the whole, such options remain a limited auxiliary to conventional CO$_2$ producing power plants. With some reluctance, fission reactors are now seen as the only near term power option that could meet high demands and limit CO$_2$ production. While their reliability approaches 99$\%$, their 1$\%$ failures have had catastrophic consequences, and with them necessarily comes a growing threat of nuclear proliferation. 

An attractive alternative that is only just approaching viability after decades of struggle is nuclear fusion. Of its various forms, fusion through magnetic confinement is the most mature. Since the construction of the first Tokamak in 1954, some 221 research-scale machines have been built, about 40 of which are still in operation. Combining the lessons learned from these devices, the major industrialized nations of the world are currently engaged in a joint effort to build the first prototype 1/2 GWatt Tokamak, the {\it{International Thermonuclear Experimental Reactor}} (ITER), now under construction in Cadarache France \cite{Goldston}. The overall scale of ITER is compared in Table~\ref{tab:table1} to the previous generations of tokamaks. The {\it{DIII-D}} tokamak operated by {\it{General Atomics}} (GA) in San Diego is typical of research-scale machines \cite{DIIID}, and is one of the most well instrumented. The {\it{Joint European Torus}} (JET) at the Culham Lab in Oxfordshire is the largest currently operating Tokamak and has come the closest to the {\it{break-even}} point where generated fusion power would be comparable to the input power \cite{JET}. ITER aims to cross that threshold with an educated leap into an unexplored regime. The expected plasma volume will be an order of magnitude larger than previous machines, and the torus field will be about double, requiring very large super-conducting coils. ITER will be a critical milestone and is regarded as a {\it{stepping-stone}}, with at least one additional iteration needed to reach a viable fusion power plant. While the costs of ITER are formidable \cite{ITER-costs}, the {\it{first}} of any new technology usually is, with savings expected only in subsequent stages. In the case of a fusion reactor, the plant costs are expected to scale roughly with the product of the plasma volume and the square of the central field, {\it{V$_{PL}$ $\times$ B$_c^{\tiny~2}$}}, so the large field volume becomes a critical factor. Furthermore, simulations for ITER have projected a drop in power and efficiency by almost a factor of two from only a 10$\%$ reduction in field \cite{pach-08}, and this has raised a specter of the consequences of possible neutron degradation to the super-conducting coils. For all of these reasons, any methods that increase the fusion power without changing the field will have a significant impact. Here we explore the option of spin-polarizing the fuel.

\begin{table}
\caption{\label{tab:table1}Tokamak characteristics of plasma volume, central field, fusion power, and net efficiency {\it{Q}}, defined as the ratio of power from fusion reactions to auxiliary (input) power. The torus coils are either normal ({\it{NC{\tiny~}}}) or superconducting ({\it{SC{\tiny~}}}), as indicated in the last column. DIII-D is typical of research-scale machines and JET is the largest tokamak presently in operation.  ITER is under construction. }
\begin{ruledtabular}
\begin{tabular}{lrcccc}

\- & V$_{PL}$ & B$_c$ & P & Q & coils \\ 
\- & (m$^3$) & (Tesla) & (MW) & \- & \-  \\ \hline
DIII-D:    &   20 & 2.1 &    \-  &  {\footnotesize $\ll$1} & NC  \\
JET:       &   90 & 3.8  &    16  &  {\footnotesize 2/3} & NC  \\
ITER:     & 700 & 5.3  &  500  &  $\sim$10 & SC  \\ 

\end{tabular}
\end{ruledtabular}
\end{table}

\section{Polarized fusion reactions} 

The primary reaction for fusion power is D~+~T~$\rightarrow \alpha$~+~{\it{n}} . Both this, as well as its isospin-mirror process D~+~$^3$He~$\rightarrow \alpha$~+~$p$, are dominated at low energies by spin 3/2 {\it fusion} resonances that are just above the particle-decay thresholds in the compound nuclei $^5$He and $^5$Li. At keV energies such reactions are dominated by s-wave processes. Under these conditions it is obvious that a spin 1 deuteron and a spin 1/2 triton (or $^3$He) will preferentially fuse into a spin 3/2 state, when their spins are parallel, so an alignment of their spins should lead to an enhancement of the reaction cross section. 

While this had been known for decades \cite{gold-34},
it was not at all clear if spin alignments could survive in a $10^8$ K plasma for long enough to be useful. However, in 1982 Kulsrud, Furth, Valeo and Goldhaber 
\cite{KFG-82} predicted time scales for polarization loss in a plasma that were in fact much longer than the characteristic fuel burn-up period. 
Since polarization-enhanced cross sections could potentially increase efficiency, that paper led to considerable theoretical activity over the subsequent decade.

\begin{figure}
\center{
\includegraphics[scale=0.65]{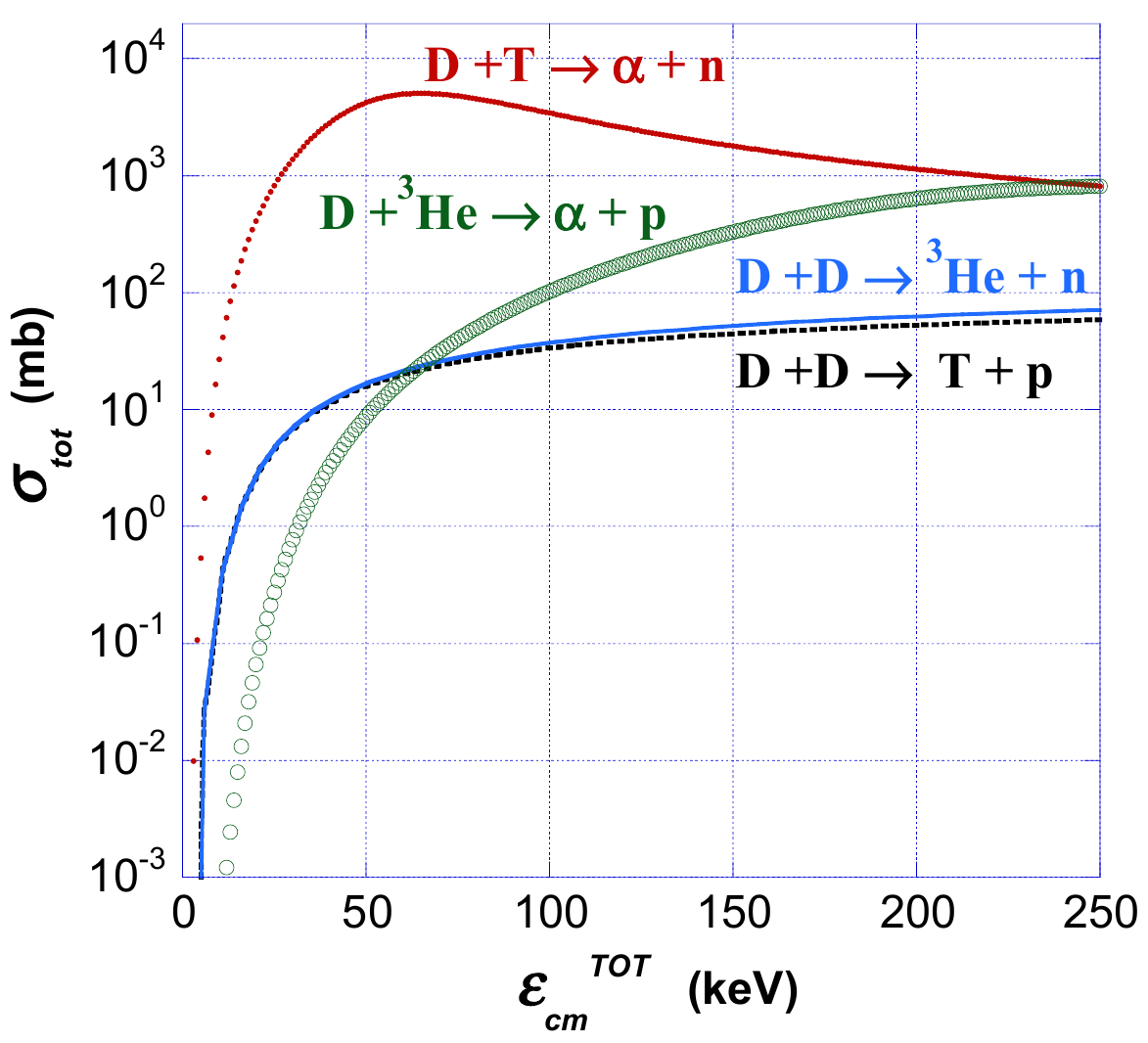}
\caption{Total cross sections for relevant fusion reactions, \cite{bosh-92}.}
\label{SIGs} }       
\end{figure}

Angle-integrated cross sections for the main fusion processes are shown in figure~\ref{SIGs} as a function of their total center of mass (CM) kinetic energy \cite{bosh-92}, assuming no polarization in the entrance channels. While the D~+~T~$\rightarrow \alpha$~+~$n$ and D~+~$^3$He~$\rightarrow \alpha$~+~$p$ reactions become comparable above 250 keV, the former completely dominates at low energies. A plasma contains a distribution of energies and the net fusion rate of two species is determined by their densities, $n_1$~ (cm$^{-3}$) and $n_2$~(cm$^{-3}$), the effective plasma volume $V$~(cm$^3$), and the cross section averaged over a Maxwell-Boltzmann velocity distribution 
$\langle \sigma v \rangle$~(cm$^3$s$^{-1}$) \cite{bah-66},
\begin{equation}
\langle \sigma v \rangle = \frac{4c}{{\sqrt{2\pi M_r}} (k_B T)^{3/2}}\int{e^{-\epsilon /k_B T} \epsilon \sigma (\epsilon) d \epsilon }.
\label{ASeq:1}
\end{equation}
Here, $\epsilon$ is the total CM kinetic energy, $M_r = \frac{m_1 m_2 }{(m_1 + m_2)}$ is the reduced mass and $k_B T$ is the ion temperature expressed in keV using Boltzmann constant.

The integral of eqn.~(\ref{ASeq:1}) is essentially saturated below about 50 keV for D~+~T, and below about 100 keV for D~+~$^3$He \cite{AMS_FUSbook}.
For context, the projected average temperature for the ITER plasma is about 12 keV, and the expected peak temperature is 18 keV \cite{aym-02} . 

The $^5$He and $^5$Li compound states in the D~+~T~$\rightarrow \alpha$~+~$n$ and D~+~$^3$He~$\rightarrow \alpha$~+~$p$ reactions are isospin-mirror nuclei. Their spin 3/2 capture ({\it{fusion}}) resonances lie just above their particle decay thresholds, at 16.84 MeV in $^5$He (50 keV above the D~+~T threshold) and at 16.87 MeV in $^5$Li (210 keV above the D~+~$^3$He threshold) \cite{til-02}. 
Both compound nuclei have a gap of over 2.3 MeV between these J=3/2 \textit{fusion} resonances and the next excited level, so that the only possible way higher states can contribute to reactions at tokamak energies is through the low-energy tails of broad states. While some excited states do in fact have quite large widths 
(\textit{e.g.} a $J_{\pi}=1/2^+$ level in $^5$Li at 20.5 MeV above the ground state with a 5 MeV width), interference effects from these tails have been examined and found to alter the polarized angular distributions by at most $2 -$to$- 3$\% for D~+~$^3$He, depending upon energy, and their effect is even smaller for D~+~T \cite{san}. All such interference effects will be neglected here. Then, to an excellent approximation the Maxwell-averaged cross section for the D~+~T~$ \rightarrow \alpha$~+~$n$ reaction simplifies to \cite{san}, 
%

\begin{eqnarray}
\langle d\sigma (\theta) v \rangle = \frac{1}{4\pi} \langle \sigma_0 v \rangle W(\theta)  
= \frac{1}{4\pi} \langle \sigma_0 v\rangle  \Big \{ 1 - \frac{1}{2} P_D^V P_{T} \ \ \ \ \ \ \ \nonumber  \\
+ \frac{1}{2} \big[ 3 P_D^V P_{T} \sin^2\theta +
\frac{1}{2} P_D^T \big(1-3\cos^2\theta\big)\big] \Big \}. \ \ \ \ \label{ASeq:2} 
\\ \nonumber
\end{eqnarray}

Here the leading factor $\langle \sigma_0 v \rangle / 4\pi$ is the isotropic rate, that would be observed in the absence of initial-state polarization. The polarization factors follow the usual definitions: $P_{T} = N_{+1/2} - N_{-1/2}\in [-1,+1]$  is the degree of triton polarization, determined by the sub-state population fractions, relative to the tokamak magnetic field direction. Similarly,   $P_D^V = N_{+1} - N_{-1}\in [-1,+1]$ is the deuteron vector polarization and $P_D^T = N_{+1} + N_{-1} - 2 N_0 \in [-2,+1]$  is the associated deuteron tensor polarization. The polar (pitch) angle $\theta$ is measured relative to the local magnetic field, and the reaction yield is symmetric in azimuthal (gyrophase) angles.
(The corresponding expression for D~+~$^3$He~$\rightarrow \alpha$~+~$p$ is identical in form, with $P_T$ replaced by $P_{^3He}$.)

Several observations are worth noting about the structure of eqn.~(\ref{ASeq:2}). The simple factorization, into an isotropic leading term determined by nuclear reduced matrix elements and a purely angular function $W(\theta)$, holds as long as we neglect the interference terms, which is in fact an excellent approximation \cite{san}. 
If the T (or $^3$He) is unpolarized, the angular dependence is modified from isotropy only by the tensor polarization of the deuteron. 
However, the angular factor of that term, $(1 - 3 \cos^2\theta)$, integrates to zero in the total cross section so that the total reaction rate is not modified. 
(Nonetheless, as discussed in \cite{kul-86,pace-16}, potentially this could provide a measure of control over the direction of neutrons from D~+~T reactions.) 
If the deuteron is unpolarized, the angular function $W(\theta)$ reduces to unity. 
Thus the total fusion reaction rate differs from the unpolarized case only if \textit{both} reacting species are polarized.

\begin{figure}[b]
\center{
\includegraphics[scale=0.6]{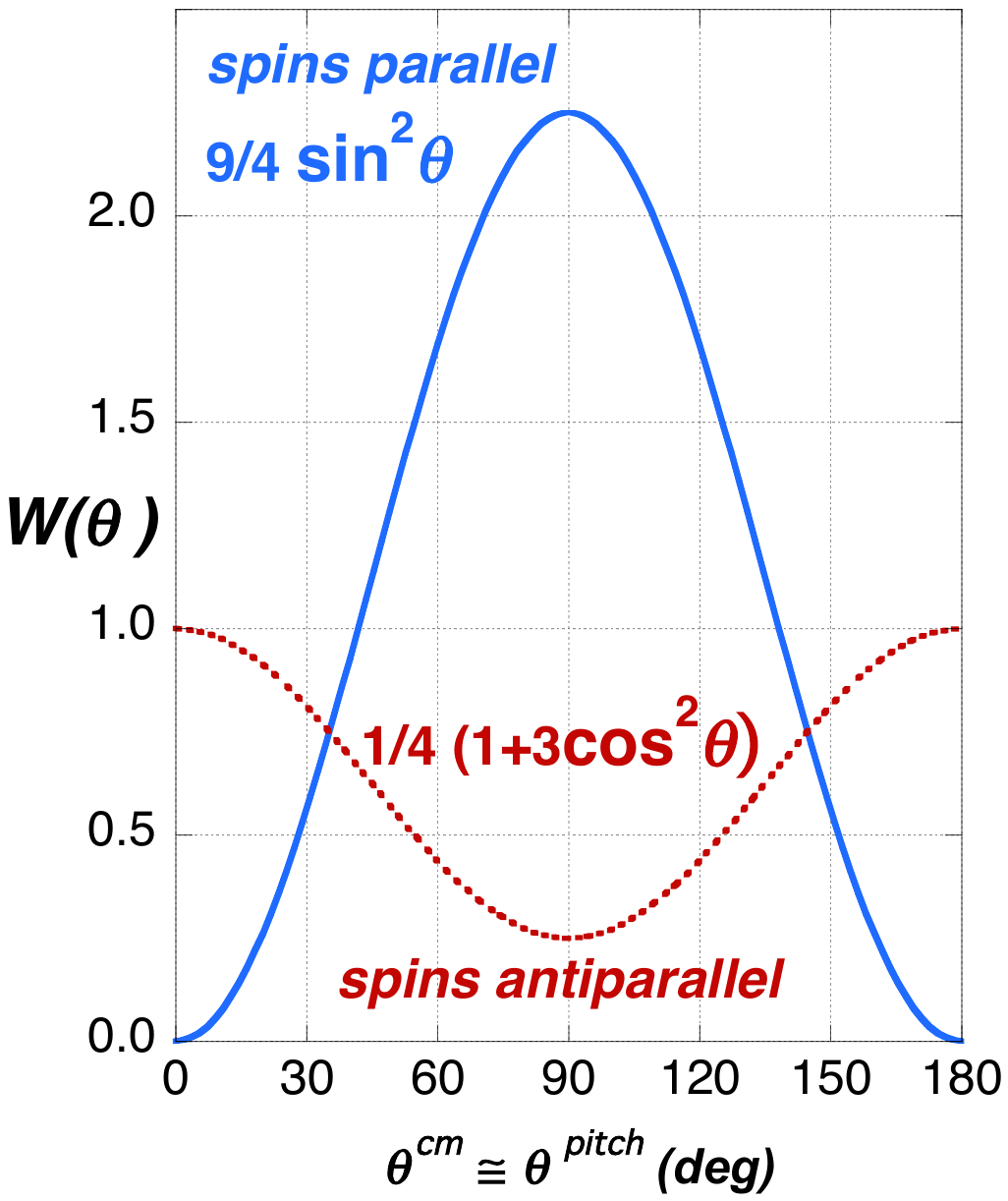}
\caption{Ions follow helical paths around the local magnetic field lines. The pitch (polar) angles $\theta$ of reaction products are measured relative to the local field direction. The solid blue curve gives the expectation for the angular factor $W$ of eqn.~(\ref{ASeq:2}) for fully polarized fuel with parallel spin alignment; the dotted red curve shows the expectation if the spins are anti-aligned.}
\label{DSGpol} }      
\end{figure}

The angular distributions calculated from eqn.~(\ref{ASeq:2}) for full vector polarization $\{|P_D^V|=1$ \big($\Rightarrow$ $P_D^T=1$\big), and $|P_{(T ~or~^3He)}|=1\}$ are plotted in 
figure~\ref{DSGpol}. 
For the case where the D and T (or $^3$He) spins are both parallel to the magnetic field, the angular function of eqn.~(\ref{ASeq:2}), $W(\theta)$, reduces to 
$9/4\ \sin^2 \theta$, which is plotted as the solid (blue) curve; for the corresponding case of anti-parallel spin alignment, $\{P_D^V = \pm 1$, $P_D^T = 1$, and 
$P_{(T ~or~^3He)} = \mp 1$, $W(\theta)$ becomes $1/4(1 + 3\cos^2\theta)$ which is shown as the dotted (red) curve. The enhancement from parallel spin alignment is obvious.

Integrating eqn.~(\ref{ASeq:2}) over all pitch $(\theta)$ and gyro-phase $(\varphi)$ angles, determines the total reaction rate as,
\begin{equation}
\langle \sigma v \rangle = \langle \sigma_0 v\rangle \big \{ 1 + \frac{1}{2} \overrightarrow{P}_D^V \cdot \overrightarrow{P}_{(T~or~^3He)} \big  \} ,
\label{ASeq:3}
\end{equation}

\noindent where here we write the polarization factors as vectors, reflecting their range between [-1, 1]. Thus, if the spins of the reacting species are anti-parallel, the reaction rate is 1/2 of the unpolarized rate. But if the initial spins are parallel, the reaction rate is enhanced by a factor of 1.5, which is the original observation of Kulsrud {\it{et al}} \cite{KFG-82}.
\section{Enhancements in Large scale machines} 
 
The fields and dimensions of high power tokamaks such as ITER are designed to confine the alpha fusion products. Coulomb interactions of these alphas with electrons and with fuel ions raise the plasma temperature. At the expected ITER plasma energies of 12-18 keV \cite{aym-02}, $\langle \sigma v \rangle/T_{ion}^2$ is approximately constant \cite{AMS_FUSbook}.
This quadratic increase of the fusion rate $\langle \sigma v \rangle$ with ion temperature has two ramifications. First, it leads to an additional non-linear increase in the fusion power with polarized fuels beyond the simple factors of eqn.~(\ref{ASeq:3}). (Polarization increases the alpha yield, which increases the temperature, so that the reaction rate climbs further up the low-energy tail of the fusion resonance.) We have simulated the effects of polarized fuel in projected ITER plasmas and find power increases up to a factor of 1.75 \cite{sterling, san}.
  
In the temperature range for which $\langle \sigma v \rangle/T_{ion}^2$ is approximately constant, 
simple estimates for the fusion rate also become proportional to the forth power of the tokamak field \cite{AMS_FUSbook}. Detailed simulations for the ITER plasma show a field dependence closer to {\it{B${\tiny ~}^6$}} \cite{pach-08}. As a result, fuel polarization has the potential to compensate for $\sim$10$\%$ reductions in torus field, which could mitigate field degradation from the neutron fluence at the super-conducting coils. For future tokamaks, fuel polarization could reduce plant costs by about 20\%, which represents potentially huge savings.

\section{Depolarization mechanisms in large and research-scale tokamaks}

To be useful, fuel polarization must survive while the polarized species remain in the plasma. A variety of possible depolarization mechanisms have been investigated theoretically. A summary of past work is given in \cite{kul-86}  and the papers sited therein; recently, the issues have been revisited by Gatto \cite{gatto}. There are essentially two mechanisms of concern that survive scrutiny, interactions with the tokamak walls and resonant interactions with plasma waves. The impact of these in a large-scale (\textit{e.g.} ITER) machine versus a small research tokamak can be very different. 

Following injection, a small fraction of the fuel mass undergoes fusion in the tokamak core, while most of the ions leave the plasma without undergoing a nuclear interaction. Upon reaching the walls, these ions pickup electrons and are neutralized. At the walls, there are several potential mechanisms that can, depending upon the structure and conditions of the wall material, lead to depolarization. If these atoms subsequently reenter the plasma they can dilute its polarization. However, the consequences of wall depolarization are significantly different between a high power reactor such as ITER and the current generation of lower-power research machines. ITER cannot be fueled by external gas jets (\textit{gas puffing}) but must be fueled by pellets injected into the plasma core, since the region outside the last closed field line (the {\it{Scrape-Off Layer}}) is expected to be almost opaque to neutrals from the walls \cite{pach-08,gar-12}. In ITER, particles leaving the plasma will be swept to the diverter by convection so that the recycling of fuel from the walls, and hence the dilution of the polarization in the core, is expected to be essentially insignificant.

This is not the case in a lower-power machine, such as the DIII-D tokamak in San Diego. Potential wall-depolarization mechanisms have been discussed extensively in \cite{gre-84}, where low-Z, non-metallic materials were expected to be optimal. 
Fortuitously, the graphite walls of some research tokamaks, and DIII-D in particular, are well suited. Carbon has no conduction band, so that hyperfine interactions with polarized material are eliminated. However, the material is porous and excessive dwell times at the wall could compound the chance of encountering paramagnetic impurities. But this can be mitigated by the deposition of a thin (100 nm) layer of boron on the walls, which has been shown to dramatically increase confinement times \cite{jac-91,jac-92}.
The reduced dwell-time on a \textit{Boronized} wall, coupled with modest energy confinement times in research machines such as DIII-D ($\sim 0.2$ s), is expected to be effective in keeping wall depolarization at a minimal level for a program of spin-polarized fusion studies.

The electrons and ions of the plasma current generate electromagnetic waves. A particular class, the Alfv\'en eigenmode, arises from the periodic boundary conditions of the tokamak geometry (see \cite{pace-15} for a recent overview.)  
When an ion's orbit is in phase with the eigenmode, their interaction can result in a large displacement of the ion orbit, causing it to experience large fluctuations in magnetic field, which could cause depolarization. As discussed in several papers \cite{kul-86,coppi-86}, excitation and amplification of these collective modes can be enhanced by the anisotropic decay angular distributions of Fig.~\ref{DSGpol}.
These early studies examined the interaction between collective Alfv\'en modes and the alpha particles decaying preferentially perpendicular to the local field following polarized D+T fusion. They concluded that, while modes resonant with the deuteron spin were unlikely to be excited, depolarization times for tritium could be shortened. Since they assumed many wall-recycling times, they concluded that triton depolarization could be quite significant. 

As discussed above, recycling in an ITER-scale tokamak is not expected to be significant, which immediately limits any deleterious effects of the coupling between alpha decay angular distributions and Alfv\'en modes. Furthermore, in practice the mode properties of a plasma are highly variable, and one could contemplate developing a plasma in which specific modes are suppressed, albeit with effort. On the other hand, research machines, such as the DIII-D tokamak, are too small to drive appreciable Alfv\'en modes from charged fusion products because they are quickly lost due to their relatively large cyclotron orbits. Thus, such resonant depolarization effects are not expected to pose a fundamental limit to either a demonstration experiment on a research-scale machine or ultimate utilization in a large-scale power reactor.

\section{Proof of Principle tests in a research tokamak}
The potential benefits of fuel polarization rely on the assumption that the polarization will survive in the plasma for periods at least comparable to the energy confinement time. This must be tested before substantial research into polarized fueling scenarios is justified.  A Jefferson Lab (JLab), DIII-D/GA and University of Virginia (UVa) collaboration is preparing a demonstration test of polarization survival in the plasma of the DIII-D tokamak \cite{san}. Here, we sketch the measurement strategy that is under development to test {\it{Spin Polarized Fusion}} (SPF).

The use of tritium in research tokamaks is extremely limited. (In fact, only the JET facility has such authorization.) In addition, polarized tritium is not yet available and requires significant R\&D. Nevertheless, D~+~$^3$He~$\rightarrow \alpha$~+~$p$ can be used equally well as a test reaction, with separate pellets of polarized D and polarized $^3$He injected into a high-performance H-mode hydrogen plasma \cite{QHmode}, with temperatures comparable to the projected ITER plasma (T$_{ion}$ =$13 - 15$ keV). The resulting energetic protons will have large gyro-radii and will rapidly leave the plasma and be detected at several wall locations. Suppressing details, there are three main components to the experiment.

(1) Existing JLab facilities in Virginia can be used to diffuse about 400~bar of molecular HD into 4 mm $\O$ Gas-Discharge-Polymer (GDP) shells, supplied by GA \cite{nik-02}. These will be cooled to a solid and transferred to a dilution refrigerator + superconducting magnet system, where they will be polarized at about 12~mK and 15~Tesla, 
using the techniques discussed in \cite{bass-14}. After the spins have become frozen, they will be cold transferred to another cryostat, where RF will be used to increase the deuteron polarization by transferring H spin to D \cite{wei-13}. A deuteron polarization of about 40\% is expected, with a polarization-decay time in excess of a year. 
These pellets can then be shipped in a suitable cryostat to San Diego, loaded into a 2~K cryogenic pellet injector and fired into the DIII-D tokamak with a cold, 
supersonic hydrogen gas jet \cite{bay-07}. Apart from the filling of a thin-walled GDP shell with high pressure HD, this stage just amounts to creating a small Nuclear Physics (NP) target with standard technology. 

\begin{figure}[t]
\center{
\includegraphics[scale=0.6]{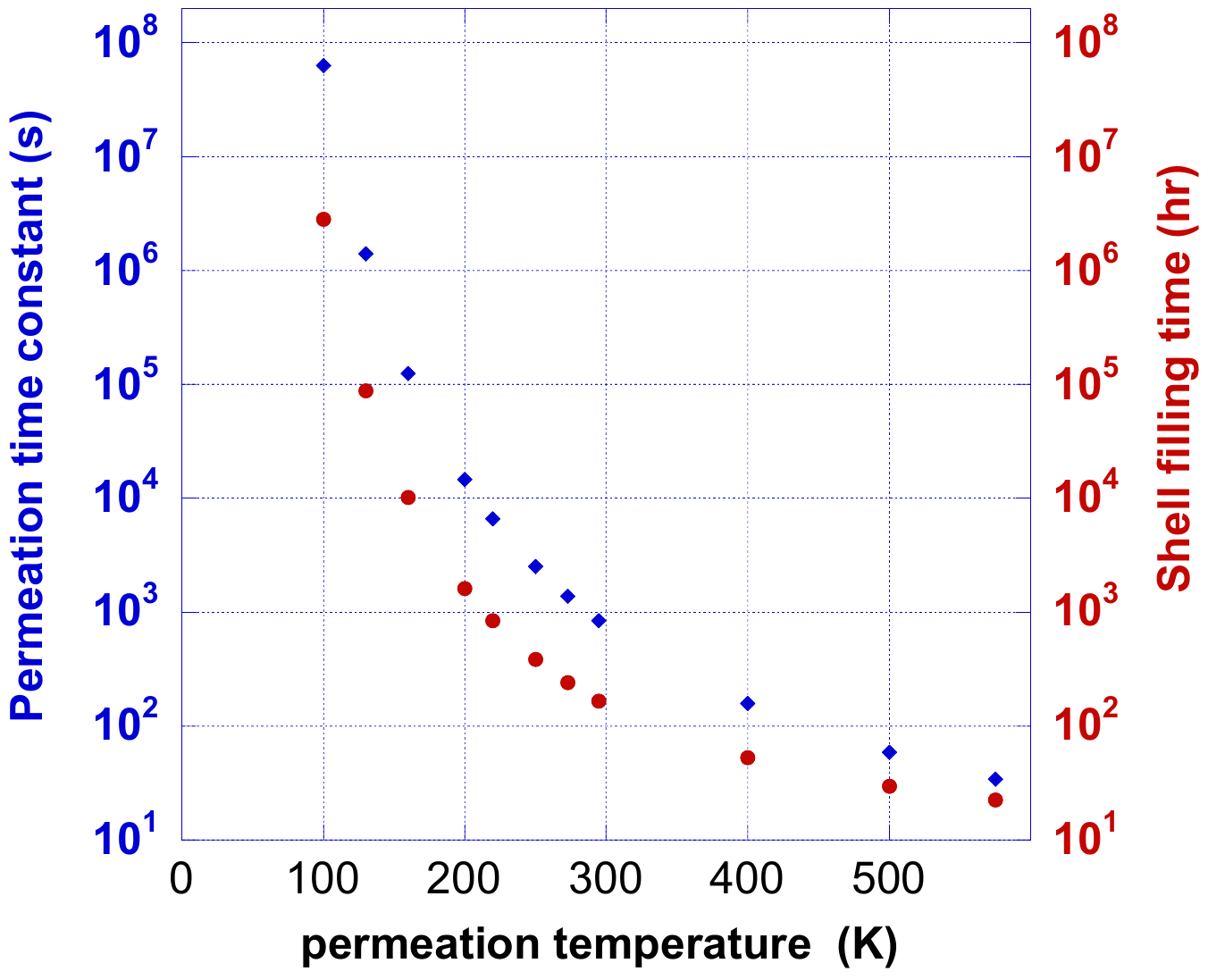}
\caption{The permeation time constant of a 4 mm \O\ x\ 20 $\mu$m wall GDP pellet as a function of temperature is shown in blue (diamond points) with its scale on the left. The time required to fill this pellet to 0.3~mmol of HD, using the sequence described in the text is plotted in red (circle points) with its scale on the right.}
\label{GDPperm}  }    
\end{figure}

The GDP shell material, C$_2$(CH$_3$)H$_2$, is commonly used for Inertial Confinement Fusion (ICF) experiments. When injecting these into a tokamak it is necessary to keep the shell material at a minimum, so that the large ionization energy of the carbon atoms does not become a sink for power and quench the plasma. Fortunately, techniques for filling thin-walled shells have already been developed and are now standard in ICF applications. One starts at an elevated temperature such as 300 $^\circ$C and increases the pressure in steps. For example, a 4 mm \O ~GDP shell with an 0.020 mm wall thickness has a permeation time constant at 300 $^\circ$C (575 K) of 34 s. The buckling pressure of the wall is 1.3~bar at this temperature. One increases the pressure of the HD gas outside the pellet in steps of 2/3 of the buckling pressure and waits 5 permeation time constants (170 s) for the pressure across the pellet wall to equilibrate. The pressure outside the pellet is then increased and the sequence repeated. The calculated time to complete this sequence and fill a 4 mm \O~GDP pellet with 0.3~mmol of HD is shown as red circles in figure~\ref{GDPperm}, 23 hours to fill to 420~bar at 575~K. Once filled, the permeation chamber is cooled cryogenically. At 100~K the pressure has reduced to 73~bar and the permeation time constant of the pellet wall is $\sim$ a year, as shown by the blue diamonds in figure~\ref{GDPperm}. At this point the pellet is effectively sealed. Cooling continues to 18~K where the pressure is 1/4~bar. At this point the HD outside the pellet is pumped away and replaced with a few mbar of $^4$He to maintain thermal contact with the walls of the permeation chamber. As the temperature is lowered past the HD triple point of 16 K, the gas inside the pellet solidifies and a routine NP target sequence begins \cite{bass-14}.

(2) The second component of the demonstration experiment requires the filling of GDP shells with polarized $^3$He. Hybrid spin-exchange optical pumping is commonly used in NP and in medical imaging experiments to create highly polarized $^3$He \cite{bab-03}. With this technique, a glass cell containing pressurized $^3$He and small amounts of rubidium and potassium ($\sim$ $10^{14}$cm$^{-3}$) are heated to over 
200$^\circ$C to vaporize the alkalis. The Rb vapor is polarized with 795 nm circularly polarized light from diode lasers, and collisions between the alkali atoms and helium transfer polarization to the $^3$He, usually in the sequence Rb~$\rightarrow$~ K~$\rightarrow$~$^3$He. After about 15 hours the $^3$He polarization has saturated at about 65\% 
and the temperature of the polarizing cell is lowered \cite{Moo-09}. The vapor pressures of Rb and K drop rapidly with temperature, so that at 20 $^\circ$C their concentrations are negligible. At this point, the $^3$He can be extracted from the polarizing cell and used to fill a GDP pellet, with the same general procedures described above in the filling of HD shells. The key difference here is that the $^3$He must be polarized first, its polarization must survive permeation of the GDP wall, and the polarization decay time within the pellet must be sufficient to allow for transfer to a pellet gun and injection into the plasma.

\begin{figure}
\center{
\includegraphics[scale=0.49]{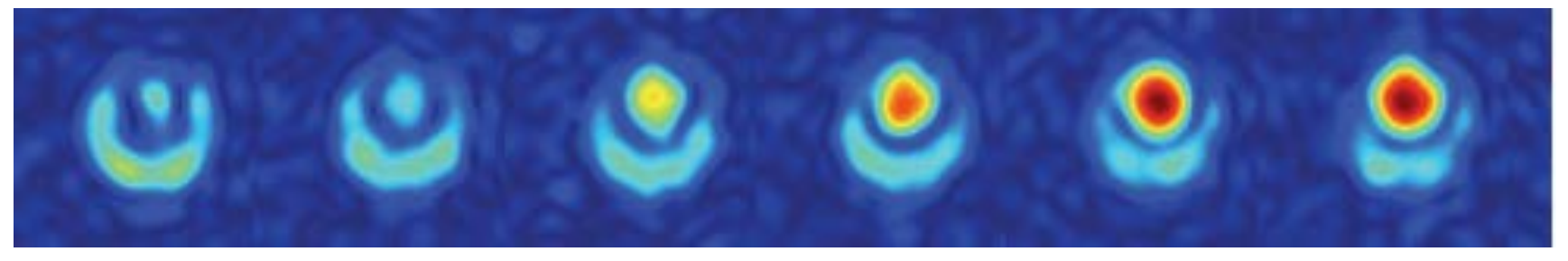}
\caption{A time sequence of MRI images showing a cross-sectional slice through a 2 mm diameter spherical GDP pellet contained in a 3 mm diameter glass tube during permeation by polarized $^3$He. Time increases to the right and the color change from blue to red indicates increasing polarization.}
\label{MRI}  }    
\end{figure}

The permeation and polarization properties of $^3$He in GDP shells are now actively being studied by a UVa-JLab team, 
using 2~mm $\O$ GDP pellets supplied by GA \cite{far}. Radiology research facilities at the UVa School of Medicine \cite{Moo-09} are being used to track the filling process by generating 3D polarization images of GDP shells during permeation, using a 1.5 Tesla commercial MRI scanner. A sample of results is shown in the time-sequenced MRI images of figure~\ref{MRI} \cite{miller}, with time increasing to the right. The intensity in the region of the pellet increases as the density of polarized $^3$He grows. A preliminary analysis of the data indicates that at least 2/3 of the polarization survives the permeation process \cite{miller}. There are several ways to improve this yet further, which are now under study.  Nonetheless, as discussed below, a SPF demonstration experiment using the existing GDP pellets is already viable.

From our initial measurements, the lifetime of the $^3$He polarization within 2 mm $\O$ GDP shells is about 5 hours at liquid nitrogen temperatures. From previous work on NP targets, we anticipate that the depolarization is dominated by interactions between the $^3$He and the pellet wall. For a SPF demonstration experiment at DIII-D we envision using 4~-~8~mm $\O$ GDP shells filled to 
about 20~bar. Thus, if the $^3$He polarization lifetime scales with the volume/surface ratio as expected from NP experience, polarization lifetimes of as much as a day could be realized. These times scales require a polarizer on site at the tokamak, and the development of dedicated equipment optimized for this purpose is planned for UVa. With such a scenario, even five hours is already much more than needed to fill pellets, transfer to a 77~K cryogenic pellet injector and fire them into the tokamak.

(3) The final component of the planned SPF demonstration experiment is the synchronized injection (to $\sim$ a few ms) of polarized D, as HD pellets from a 2~K cryo-gun, and polarized $^3$He pellets, from a 77~K cryo-gun, into DIII-D. The polarizations must be maintained by a guiding magnetic field (typically less than a kilo-gauss) throughout their flight path to the outer edge of the Tokamak. Both species are in frozen-spin configurations, with fixed sub-state populations. While injection velocities are typically $\sim$10$^3$ m/s, this and the inevitable tumbling motion down the guide tube are orders of magnitude slower than the Larmor frequencies of either D or $^3$He. As a result, the D and $^3$He spin vectors will simply follow the local field. Once in the hydrogen plasma, the spins will align along the local magnetic field. The anti-parallel configuration can be prepared using an RF transition (an {\it{Adiabatic Fast Passage}}) to flip the sub-state population so that the spin of the $^3$He (or the D, but not both) is aligned against the local magnetic field. 

The signal of spin survival consists in comparing the proton yields from successive plasma shots, in which D and $^3$He are injected with their spins alternatively parallel and anti-parallel. Assuming the anticipated polarizations of 40\% D and 65\% $^3$He, the simple expectation from eqn.~(\ref{ASeq:3}), ignoring the DIII-D acceptance (the efficiency for protons to reach specific detectors at the wall from different locations within the plasma), is
\begin{equation}
\frac{\langle \sigma^{par} v \rangle}{\langle \sigma^{anti} v \rangle}=1.30.
\label{SPFsignal}
\end{equation}

Detailed tracking simulations for DIII-D have been carried out, starting with (unpolarized) D+D plasma fusion density profiles, measured following deuteron pellet injection (DIII-D shot 96369). These fusion rates were scaled by the ratio of D+$^3$He to D+D cross sections, and then scaled again to the cross sections corresponding to the T$_{ion}$ = 15 keV expected from a Quiescent H-mode plasma \cite{QHmode}. The resulting spacial fusion profile was discretized and $\alpha$+{\it{p}} events were generated at each fusion birth location along different polar (pitch) $\theta$ and azimuthal (gyrophase) $\phi$ angles relative to the local field, weighting their relative number by the polarized angular distributions of eqn.~(\ref{ASeq:2}). The orbits of the $\alpha$ and {\it{p}} fusion products were tracked through the DIII-D magnetic field until they struck the wall of the tokamak. (Further details are discussed in \cite{sterling}.)

\begin{figure}[t]
\center{
\includegraphics[scale=0.6]{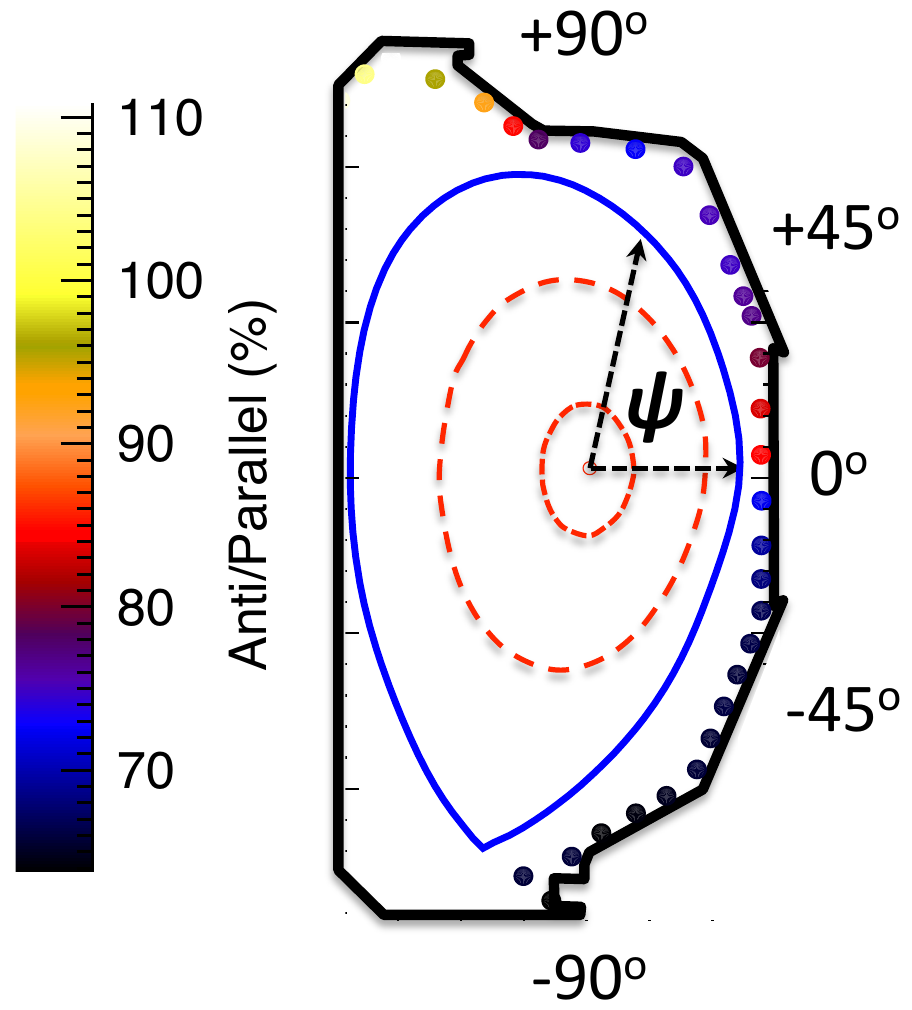}
\caption{The ratio of protons, from fusion shots with anti-parallel and parallel {\it{D}} and {\it{$^3$He}} spin alignment, as would be detected along the outer wall of the DIII-D vacuum chamber. The value of the ratio at the wall for different poloidal angles ($\psi$) is indicated by the colored circles, with their scale to the left. For physical scale, the width of the chamber (at $\psi$=0) is 1.7 m. The blue tear-drop shaped curve indicates the boundary of the {\it{Scrape-Off Layer}}, the last closed field line.}
\label{SPFratio}  }    
\end{figure}

The ratio of predicted proton yields from DIII-D shots with anti-parallel and parallel D and $^3$He spin alignment are shown in figure~\ref{SPFratio} for different locations along the inner wall of the DIII-D vacuum vessel. Not only is the strong signal of eqn.~(\ref{SPFsignal}) maintained over a large range of wall locations, but the simulations also show a striking characteristic signature in the dependence of the ratio on poloidal angle (measured at the wall locations perpendicular to the toroidal field). The latter will provide a sensitive monitor of systematic uncertainties in a demonstration experiment. 

The DIII-D is a research-scale tokamak with room temperature coils. Each {\it{shot}} starts with a 3 s ramp up to 2.1 Tesla, followed by a 10 s flat-field period in which 80 keV neutral beams heat the plasma and measurements are carried out, and then a 7 s ramp down. There are then 15 min between shots to allow the coils to cool. To capitalize on the expected strong signal evident in figure~\ref{SPFratio} requires successive plasma shots with reproducible characteristics. 
In a seasoned research tokamak such as DIII-D, the operational parameter space has been well traveled and a substantial collection of readily reproducible plasma shots has been documented (see \cite{pace-16} for some specific comparisons). For the present study, the important metric is the product of ion density and temperature, integrated over the plasma volume and over the confinement time. A review of past D~+~D experiments at DIII-D suggests a variability in the neutron production rate of less than 10\% for plasmas with central ion temperatures of T$_i \approx$8~keV. 

\begin{figure}[b]
\center{
\includegraphics[scale=0.55]{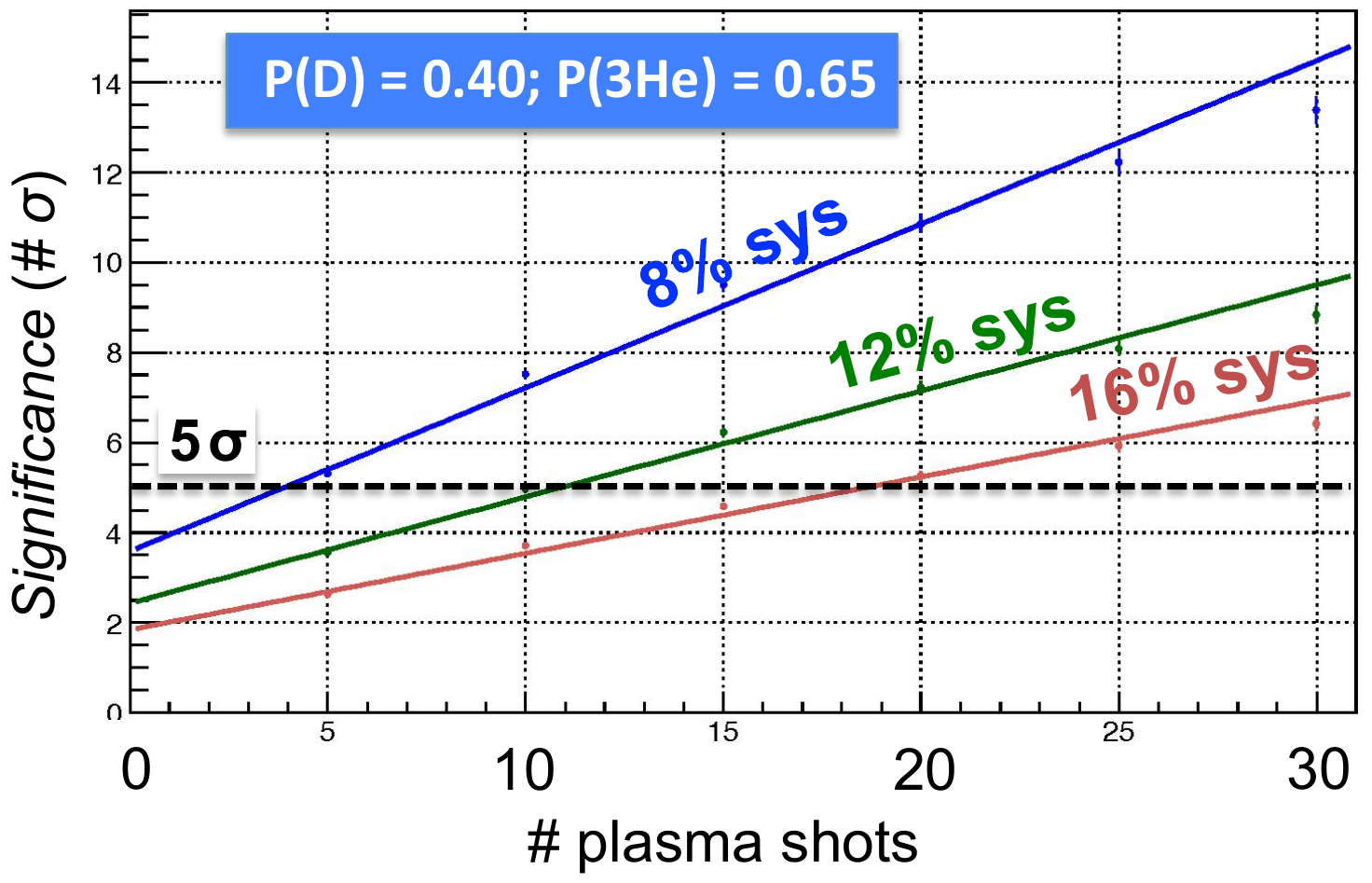}
\caption{The number of plasma shots required to reach a given confidence level in terms of the statistical significance ($\sigma$), for different assumptions on the systematic shot-to-shot variation. The 5$\sigma$ level for a definitive demonstration is indicated as the black dashed line.}
\label{shotMC}  }    
\end{figure}

To examine the impact of such shot-to-shot variations, a Monte Carlo study has been carried out. The results are summarized in figure~\ref{shotMC}. With an 8\% systematic shot-to-shot variations, a $5\sigma$ determination is obtained with merely four plasma shots in each spin orientation. A 16$\%$ systematic variation would raise the required number of shots for a $5\sigma$ result to 18 (for each spin orientation). These results assume polarizations of 40\% D and 65\% $^3$He, with the latter requiring no  loss of $^3$He polarization during permeation of the fuel pellet. Should the current 30$\%$ losses (as discussed above) not be overcome, an  8\% systematic variation would require 10 shots in each polarization orientation and 16\% systematic variation would require 30 shots for a demonstration experiment. DIII-D is capable of generating about 30 plasma shots per day, so this difference, while not fundamental, is not trivial. Work is underway to survey high-T$_{ion}$ plasmas in DIII-D and empirically determine their reproducibility level.
\\
\section{Backgrounds and detection strategy}
\label{ASsec:6}
In a D~+~$^3$He~$\rightarrow \alpha$~+~$p$ polarization survival demonstration experiment, a hydrogen plasma would be used to avoid diluting the spins of the injected reactants. Nonetheless, a chain of parallel and secondary reactions can also lead to the production of alphas and protons. These are listed with their associated energy release ($Q$ values) in eqn.~(\ref{ASeq:6}). Note that while triton production, followed by a subsequent D~+~T reaction can generate an alpha of comparable energy to the 
D~+~$^3$He~$\rightarrow \alpha$~+~$p$ fusion channel, the only secondary protons that are produced are low in energy and easily distinguished from those of the initial 
D~+~$^3$He~$\rightarrow \alpha$~+~$p$ process. It is for that reason that  proton detection is preferred.	

\begin{eqnarray}
^3\textrm{He~+~D}\Rightarrow &\alpha\textrm{~+~p}& ~(Q=+18.3\textrm{~MeV}) \label{ASeq:6}\nonumber\\
&&\nonumber \\
\textrm{D~+~D}\Rightarrow &^3\textrm{He~+~n}&~(Q=+3.3\textrm{~MeV}) \\
\textrm{D~+~D}\Rightarrow &\textrm{T~+~p} &~(Q=+4.0\textrm{~MeV}) \nonumber\\
\textrm{D~+~T}\Rightarrow  &\alpha\textrm{~+~n}&~(Q=+17.6\textrm{~MeV}) \nonumber
\end{eqnarray}

In principle, a dilution of the signal of interest can come from a two-step burn-up process, in which an initial reaction of two polarized deuterons produces a $^3$He,
following the second of the reactions listed in eqn.~(\ref{ASeq:6}). This resulting $^3$He is born with 825~keV in the lab frame and can fuse with another deuteron. The associated CM total energy for such a $^3$He~+~D secondary reaction is 330~keV, which puts it just above the peak in the fusion cross section for this channel (figure~\ref{SIGs}). This daughter $^3$He slows down through Coulomb collisions with electrons in the plasma, and in so doing crosses through the fusion resonance \cite{heid-83}. However, the rate of the primary $^3$He~+~D reaction providing the \textit{signal} is simply proportional to {\it{n(D)}} where, for a hydrogen plasma, the density {\it{n(D)}}  is just determined by the injected polarized fuel. The rate of the D~+~D reactions producing $^3$He 
(second of eqn.~(\ref{ASeq:6})) is proportional to {\it{n(D)$^2$}}, and the rate of the subsequent $^3$He~+~D burn-up process is proportional to {\it{n(D)}}. Thus, the dilution of the signal of interest by the two-step burn-up process is quenched by the factor {\it{n(D)/n(D)$^3$}}, and so is essentially negligible. 

Options for proton detection methods are under review. Both Fast Ion Loss Detectors \cite{chen-10} and Silicon surface barrier detectors \cite{heid-84} have been successfully used at the DIII-D outer wall.

\section{Summary}
In the light of recent simulations of the ITER plasma, it now seems quite likely that fuel polarization could have a major impact on performance. While there have been significant developments in polarizing materials in the last several decades, these have been driven by nuclear physics and medical imaging with very different goals in mind ({\it{eg.}} few mole samples with lifetimes of 10$^5$ to 10$^8$ s, whereas kmoles per day would be needed to fuel a power reactor, but with mere $\sim$10 s lifetimes). Since every characteristic of polarization comes at a cost, significant R\&D will be required to develop polarized fuel that is tailored to fusion requirements. However, prior to such an investment polarization survival in a plasma must be verified, and existing NP and medical imaging techniques can be used for such a demonstration in a cost effective way.  

Work is underway to take such a polarization survival experiment to the next stage of readiness. Tests are ongoing to maximize the $^3$He polarization that can be permeated into polymer pellets, which can then be injected into the tokamak. DIII-D shot variation studies are planned for high-T$_{ion}$ Quiescent H-mode plasmas, and preliminary designs of cryogenic injectors for polarized pellets are under consideration. 

Finally, we note that a European {\it{PolFusion}} collaboration based at Forschungszentrum J\"ulich is pursuing work that could potentially shed light on polarized D+D reactions in a plasma. The polarized cross sections for such reactions are considerably more complicated than for D + $^3$He, and at low energies the D+D nuclear matrix elements are essentially unknown. The {\it{PolFusion}} group are preparing experiments with a polarized molecular beam and polarized gas target \cite{Ciullo,Toporkov}. Depending upon their findings, these could provide an interesting parallel avenue for SPF studies.

We would like to thank L. Baylor of ORNL and W. Heidbrink of U. California-Irvine for fruitful discussions. This work was supported by the U.S. Department of Energy, Office of Science, Office of Nuclear Physics under contract DE-AC05-06OR23177, by General Atomics Internal Research and Development Funds and by a grant from the University of Virginia Research and Initiative Fund.

\end{document}